\documentclass[letterpaper%
,prl%
,twocolumn%
,notitlepage%
,superscriptaddress%
,showpacs%
]{revtex4-1}

\PassOptionsToPackage{colorlinks}{hyperref}

\usepackage {amsmath}
\usepackage{graphicx}
\usepackage{braket}
\usepackage{amsthm}
\usepackage{amssymb}
\usepackage{url}
\usepackage{color}
\usepackage{mathtools}

\usepackage{bm}

\usepackage{tikz}
\usetikzlibrary{decorations}
\usetikzlibrary{patterns,snakes}

\newcommand{\poly}{\text{p}}

\def\op#1{\hat{#1}}

\def\id{I}

\def\1{\mat{\id}}

\def\mat#1{\bm{\mathrm{#1}}}


\usepackage{hyperref}
\graphicspath{ {./Figures/} }

\begin{document}
\title{Measurement-Based Linear Optics}
\author{Rafael N. Alexander} \email[Email: ]{rafael.alexander@sydney.edu.au}\affiliation{School of Physics, The University of Sydney, NSW, 2006, Australia} \affiliation{School of Science, RMIT University, Melbourne, VIC 3001, Australia}
\author{Natasha C. Gabay} \affiliation{School of Physics, The University of Sydney, NSW, 2006, Australia}
 \author{Peter P. Rohde}%
\affiliation{Centre for Quantum Software \& Information (QSI), Faculty of Engineering \& Information Technology, University of Technology Sydney, Sydney, NSW 2007, Australia}
 \author{Nicolas C. Menicucci} \email[Email: ]{ncmenicucci@gmail.com}\affiliation{School of Physics, The University of Sydney, NSW, 2006, Australia} \affiliation{School of Science, RMIT University, Melbourne, VIC 3001, Australia}

\date{\today}
\begin{abstract}
A major challenge in optical quantum processing is implementing 
large, stable interferometers.
We
offer
a novel approach: virtual, measurement-based interferometers that are 
programmed on the fly solely by the choice of homodyne measurement angles. 
The effects of finite squeezing 
are
captured as uniform amplitude damping.
We compare our proposal to existing (physical) interferometers and consider its performance %
for \textsc{BosonSampling}, which could demonstrate post-classical computational power in the near future.
We prove its efficiency in time and squeezing (energy) in this setting.
\end{abstract}
\maketitle
%
%
\emph{Introduction.}---%
%
%
%
%
Large-scale stable interferometers form the backbone of any optical architecture for processing photonic quantum information. 
This includes schemes for universal quantum computation, including linear-optics quantum computing~\cite{Knill2001}, continuous-variable~(CV) hybrid quantum computing~\cite{vanLoock2008,Andersen2015}, and atomic-optical hybrid schemes~\cite{Kim2013}, as well as other applications such as \textsc{BosonSampling}~\cite{Aaronson2011}, quantum metrology~\cite{Motes2015}, quantum walks~\cite{Rohde2011}, and homomorphic encryption~\cite{Rohde2012}. 

In conventional experiments, these interferometric networks are typically built up out of bulk linear-optical elements (i.e.,~beamsplitters and phase delays)%
. %
Though relatively simple to implement, these networks are limited by the scale and complexity afforded by the laboratory optics bench and are therefore unsuitable for large-scale applications~\cite{Li2015}.
One approach to solving this problem is to leverage integrated optics technology. Miniaturized optical elements can be lithographically printed on a chip, which enables far greater scalability~\cite{Politi2008, Carolan2015}. Though this approach has shown great potential, such experimental architectures still fall short of the required scale for useful applications by several orders of magnitude%
. 

Here we take a different path to large-scale and compact linear optics: \emph{measurement-based linear optics}~(MBLO). Rather than passing physical modes through optical elements in real space, MBLO implements very large \emph{virtual} interferometers using highly compact cluster-state machinery~\cite{Menicucci2011a, Wang2014, Yokoyama2013, Chen2014, Yoshikawa2016}.

Though any universal CV cluster state can in principle implement linear optics by measurement-based quantum computation (MBQC)~\cite{Gu2009}, here we focus on a particular resource state called the \emph{quad-rail lattice} (QRL)~\cite{Menicucci2011a, Wang2014, Alexander2016} for three reasons.  
  First, it can be generated on an unprecedented scale from compact experimental setups~\cite{Yokoyama2013,Chen2014, Yoshikawa2016} using either temporal~\cite{Menicucci2011a} or frequency modes~\cite{Wang2014}. Second, single- and two-mode linear-optics gates are naturally implemented on this state by using a recently introduced measurement protocol~\cite{Alexander2016} that minimizes noise due to finite squeezing~\cite{Alexander2014}.
Finally, this noise---which is ubiquitous in CV MBQC and usually appears as additive Gaussian noise~\cite{Gu2009,Menicucci2014,Alexander2014}---can be coaxed into appearing as pure photon loss with efficiency ${\gamma = \tanh^2 r}$ for each simulated optical element, where $r$ is the overall squeezing parameter of the QRL. %

As an application, we discuss efficient \textsc{BosonSampling}~\cite{Aaronson2011} using MBLO. Demonstrating post-classical computing with \textsc{BosonSampling} lends itself naturally to MBLO because of the size and variability of the interferometer required%
. We prove that \textsc{BosonSampling} 
using MBLO is simultaneously efficient in time and squeezing (as measured by average energy~\cite{Liu2016}). Such efficiency is necessary to show post-classical processing power.
%

\begin{figure}[t]
\includegraphics[width=0.8\linewidth]{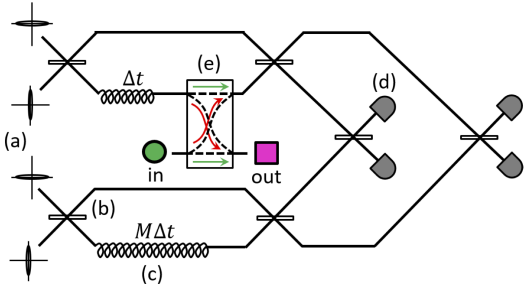}
\caption{Generating and measuring the temporal-mode quad-rail lattice~\cite{Menicucci2011a}. Squeezed vacuum states~(a) pass through six balanced beamsplitters~(b) and two delay loops~(c) with delays $\Delta t$ and $M \Delta t$, with $M$ an odd integer. Cluster modes are generated and measured with homodyne detection~(d) at the same rate. After $ML$ time steps of duration~$\Delta t$, the QRL is an ${(M\times L)}$-macronode lattice. Input (output) states can be inserted (removed) using a switching device~(e)~\cite{Yokoyama2013} on any of the four rails (one example shown).
}
\label{fig:QRLgen}
\end{figure}

%
\emph{Linear optics with cluster states.}---%
For concreteness, we choose to illustrate MBLO using the temporal-mode implementation of the QRL~\cite{Menicucci2011a}, although analogous results hold for the frequency-mode version~\cite{Wang2014}. The full experiment to generate and use the QRL using temporal modes is shown in Fig.~\ref{fig:QRLgen}. Its compactness is evident. Homodyne detection alone programs and implements the desired linear optics.

The QRL possesses a four-layered square-lattice graph~\cite{Menicucci2011a, Wang2014} and enables universal quantum computation by the measurement protocol proposed in Ref.~\cite{Alexander2016}. Lattice sites consist of four physical modes and are referred to as \emph{macronodes}. Computation on the QRL acts on the \emph{distributed modes} (labeled~\mbox{$a$--$d$}) within each macronode~\cite{Alexander2016}. These are linear combinations of the physical modes (labeled \mbox{1--4}):
${(\hat{a}_{a}, \hat{a}_{b}, \hat{a}_{c}, \hat{a}_{d}) \coloneqq (\hat{a}_{1}, \hat{a}_{2}, \hat{a}_{3}, \hat{a}_{4}) \mat A}$,
where $\mat A$ is a $4 \times 4$ %
 matrix that corresponds to the Heisenberg action of the last four balanced beamsplitters in Fig.~\ref{fig:QRLgen}~\cite{Alexander2016}.

%

%

Now, suppose that a pair of input states are encoded within distributed modes $f(a)$ and $f(b)$ of the same macronode (e.g., by the method shown in Fig.~\ref{fig:QRLgen}), where $f$ is an arbitrary automorphism on the mode labels $\{a, b, c, d\}$ reflecting the permutation symmetry inherent to computing on the QRL~\cite{Alexander2016}.  Then, mode~$f(c)$ and its partner~$f(c)^\prime$ (indicated by the prime) within an adjacent macronode are in a two-mode squeezed state~\footnote{Reference~\cite{Alexander2016} uses two-mode CV cluster states instead of two-mode squeezed states, but Footnote~1 in that reference explains that the two conventions are related simply by rotating all homodyne angles by~$\tfrac \pi 4$. We use the former convention because it connects this work more naturally to CV teleportation~\cite{Braunstein1998}.}, 
and similarly for $f(d)$ and $f(d)^{\prime}$.
Define the two-mode linear-optics gate%
\begin{align}
\hat{V}_{ij}(\theta, \phi)
&\coloneqq \op{R}_{i}(\theta) \op{R}_{j}(\theta) \left[\op{R}_i\left(\frac{\pi}{2}\right) \op{B}_{ij}(\phi)\op{R}_{j}\left(\frac{\pi}{2}\right) \right] \nonumber \\
&=\op{B}_{ij}^{\dagger}\left(\frac{\pi}{4}\right)\op{R}_{i}\left(2\xi_{+}\right)  \op{R}_{j}\left(2\xi_{-}\right)\op{B}_{ij}\left(\frac{\pi}{4}\right) \label{eq:MZform}
\end{align}
where $\xi_{\pm}\coloneqq \tfrac{1}{2} (\theta\pm\phi -\tfrac{\pi}{2})$, $\op{R}_{j}(\theta) \coloneqq 
e^{i \theta \op a_j^\dag \op a_j}
$ is a phase delay by $\theta$ on mode $j$, and $\op{B}_{i j}(\phi) \coloneqq 
e^{-\phi (\op a_i^\dag \op a_j - \op a_j^\dag \op a_i)}
$ is a variable beamsplitter on modes $i$ and $j$. The second line of Eq.~(\ref{eq:MZform}) is a Mach-Zehnder-type decomposition of~$\hat{V}_{ij}$. For any given macronode, there exists a choice of homodyne measurement angles that will implement~$\hat{V}_{ij}$ on the encoded information and teleport it from 
$f(a)\rightarrow f(c)^{\prime}$ and from $f(b)\rightarrow f(d)^{\prime}$~\cite{Alexander2016}, i.e., 
\begin{align}
\label{eq:MBMZ}
\ket{\psi}_{f(a)}\ket{\varphi}_{f(b)} \mapsto \hat{V}_{f(c)^{\prime}, f(d)^{\prime}} (\theta, \phi) \ket{\psi}_{f(c)^{\prime}}\ket{\varphi}_{f(d)^{\prime}}.
\end{align} 
This operation, which we refer to as a \emph{measurement-based Mach-Zehnder}~(MBMZ), is the primitive for MBLO. Importantly, it is merely the choice of homodyne angles that determines which gate is applied. %
The ubiquitous measurement-dependent displacements and finite-squeezing effects of CV MBQC~\cite{Gu2009} are discussed in the following section.
%

\begin{figure}[t]
\includegraphics[width=1\linewidth]{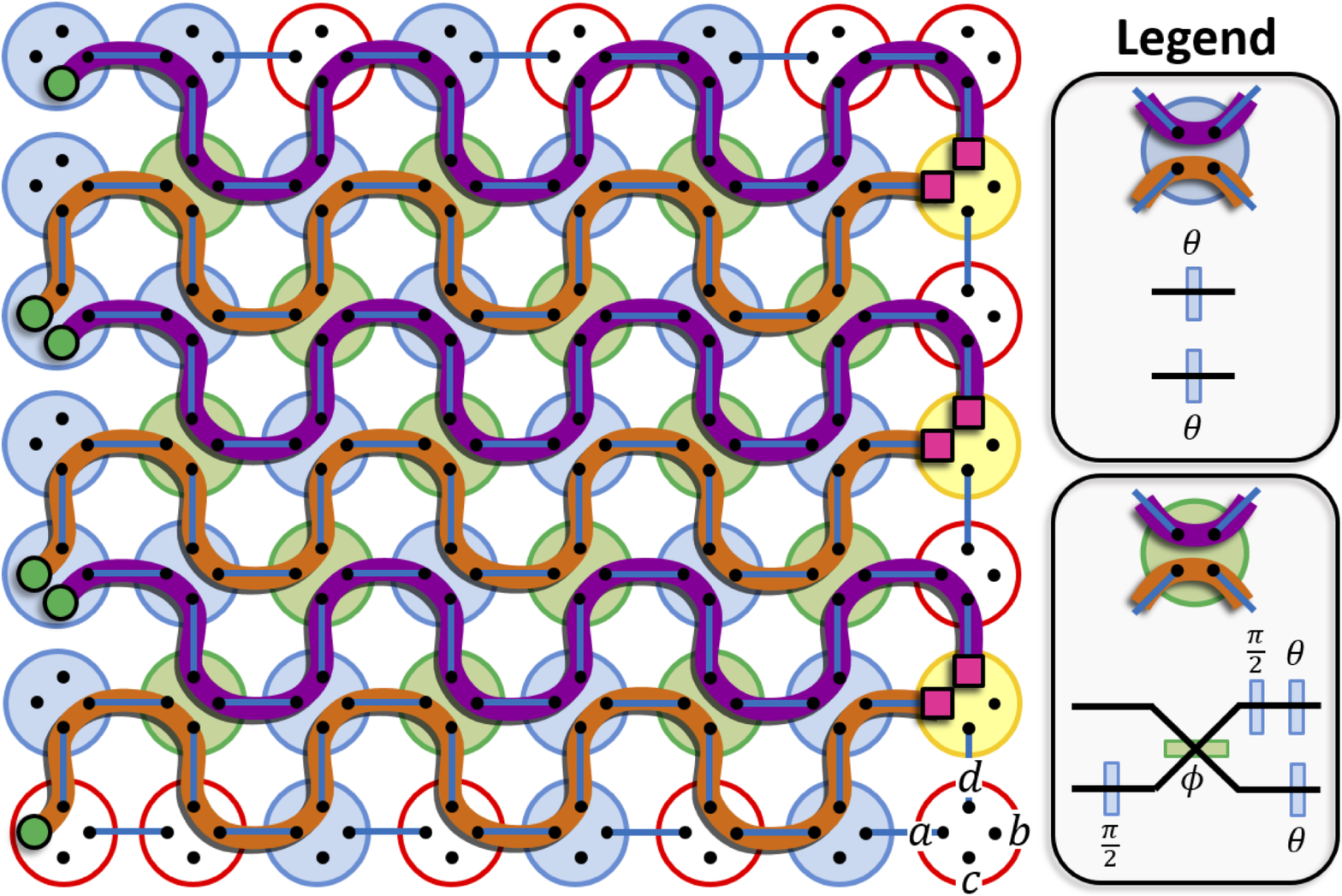}%
\caption{%
Measurement pattern of the quad-rail lattice (in terms of distributed modes~\cite{Alexander2016}) for measurement-based linear optics. The left column of macronodes contains the inputs (green circles), and the right contains the outputs (pink squares), which can be swapped in and out as shown in Fig.~\ref{fig:QRLgen}%
. Macronode columns are measured left to right, implementing measurement-based Mach-Zehnder interferometers [Eq.~\eqref{eq:MBMZ}] at each macronode (see Legend).
\label{fig:QRLnetwork}}
\end{figure}

Large networks of beamsplitters and phase delays can be implemented by composing MBMZs on the QRL. We consider a (generalizable) six-input example in Fig.~\ref{fig:QRLnetwork}. The flow of inputs (i.e.,~$f$) through each macronode is indicated by the 
orange and purple ribbons. The color of each macronode indicates which gate is applied: white applies ${\hat{V}(-\tfrac{\pi}{2}, 0)=\hat{I}}$, blue applies ${\hat{V}(\theta -\tfrac{\pi}{2}, 0) = \hat{R}(\theta)\otimes\hat{R}(\theta)}$, and green applies $\hat{V}(\theta, \phi)$. As the first two cases are single-mode gates, the inputs teleport through the macronode without interacting (like non-interacting quantum wires)~\cite{Alexander2016}.

\begin{figure}[t]
\includegraphics[width=0.8\linewidth]{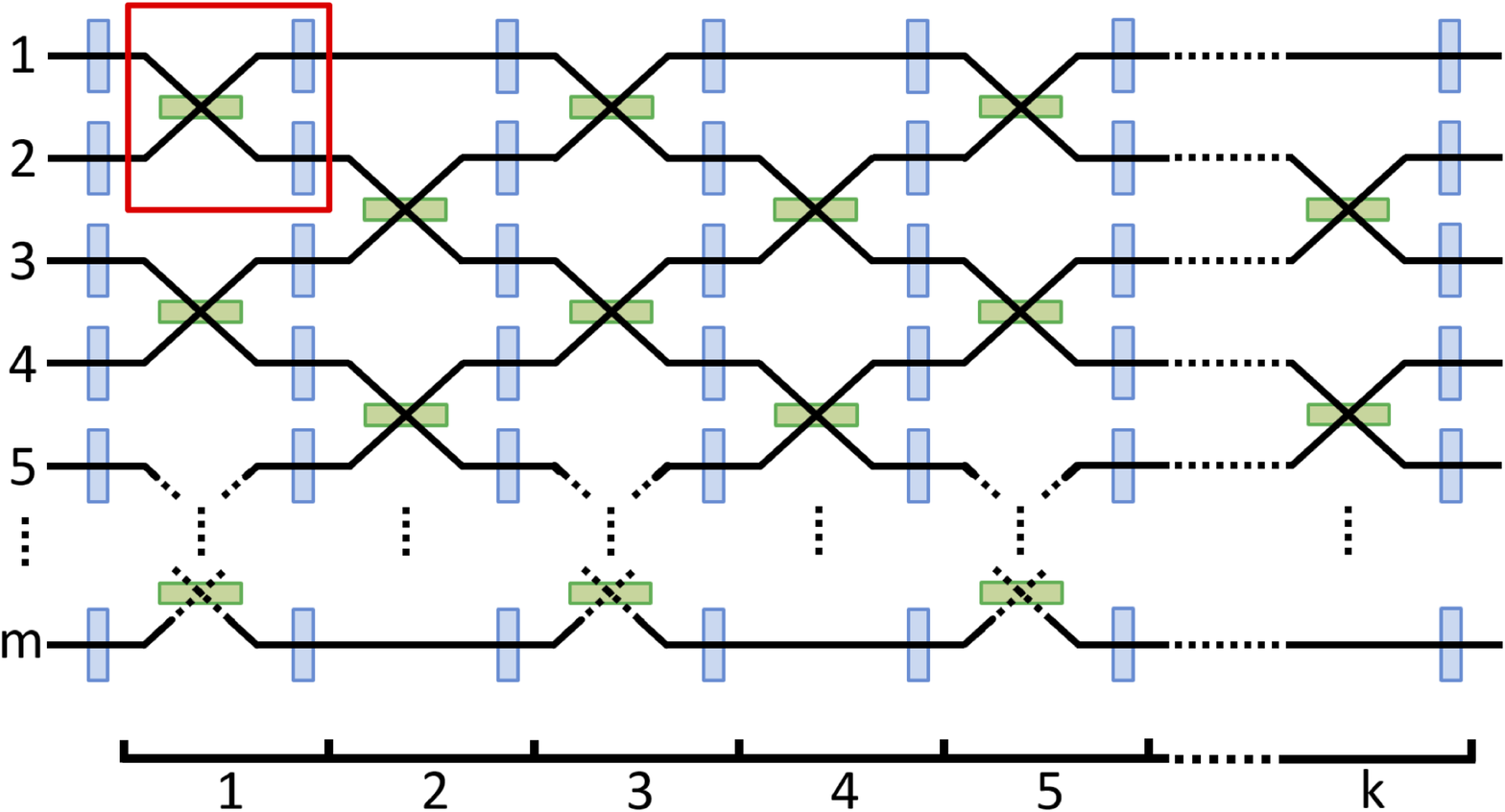}
\caption{%
The measurement pattern in Fig.~\ref{fig:QRLnetwork} is equivalent to a tessellation of the two-mode gate in the red box, which consists of 
a variable beamsplitter followed by a pair of independent phase delays.%
\label{fig:LOnetwork}}
\end{figure}

Each of the six blue macronodes in the first column of Fig.~\ref{fig:QRLnetwork} contributes one phase degree of freedom%
. Together, these are sufficient to implement arbitrary phase delays on the inputs. For columns in the bulk, green macronodes implement variable beamsplitters%
, and both green and blue macronodes contribute one phase degree of freedom each, altogether allowing arbitrary independent phase delays to act on each mode after the beamsplitters. Therefore, the total logical action of the teleportation network in Fig.~\ref{fig:QRLnetwork} is equivalent to the linear optics shown in Fig.~\ref{fig:LOnetwork}. Arbitrary $m$-mode interferometers require an $(m+1)\times(k+2)$-macronode QRL, where $k$~is the depth of the network.  A general $m$-mode interferometer can be decomposed into a depth ${k=m}$ network of this type~\cite{Clements2016}, although for some applications, a smaller network may be sufficient ($k < m$). Note that each path through the QRL crosses $2(k+1)$ macronodes (excluding the output macronodes).

\begin{figure}[t]
\includegraphics[width=.98\linewidth]{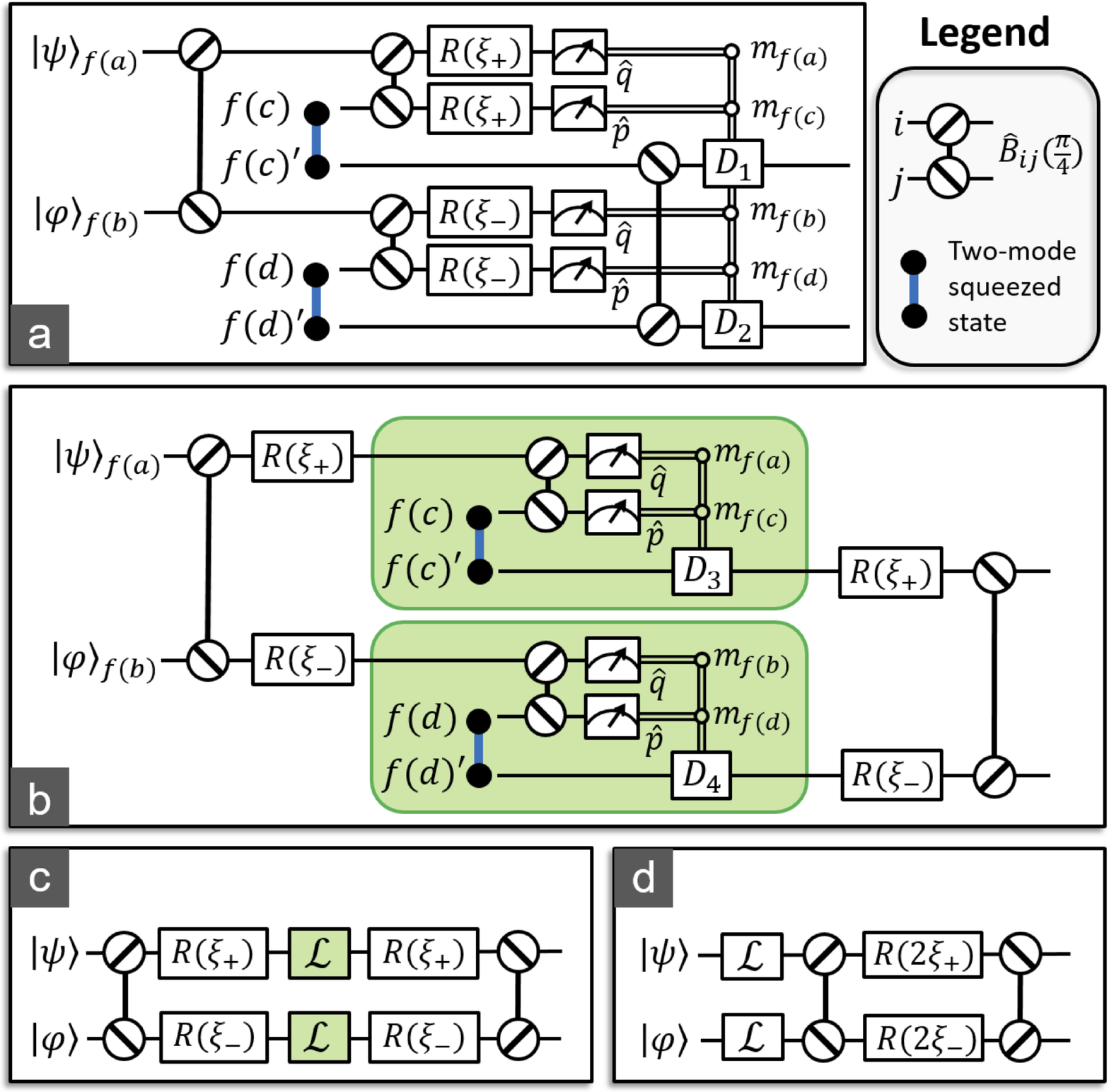}
\caption{\textbf{(a)}~Homodyne detection on one macronode in the quad-rail lattice (with respect to the distributed modes) can be represented by the above circuit~\cite{Alexander2016}. We have selected homodyne angles such that the transformation in Eq.~\eqref{eq:MBMZ} is implemented. \textbf{(b)}~The macronode circuit is decomposed as a pair of CV teleportation circuits~\cite{Braunstein1998} (shown in the green boxes) sandwiched between linear optics. 
\textbf{(c)}~Macronode measurement with finite-squeezing effects and gain-tuned displacements~\cite{Takeda2013} is equivalent to a Mach-Zehnder interferometer with equal loss~$\mathcal L$ on each arm. %
\textbf{(d)}~Because the loss~$\mathcal L$ is the same on both arms, it can be commuted to the beginning (see text).\label{fig:macro}}
\end{figure}

%
\emph{Finite squeezing as uniform loss.}---%
%
%
We now analyze the role of displacements and finite squeezing in MBLO. Each macronode measurement, illustrated in Fig.~\ref{fig:macro}(a), displaces the input states in phase space by an amount dependent on the (random) measurement outcomes~$m_{f(\cdot)}$~\cite{Alexander2016}. 
We pretend that these displacements are undone after each macronode measurement (using $\hat{D}_{1}$ and~$\hat{D}_{2}$ in the figure), but in practice all displacements will be corrected in one shot at the very end~\cite{Gu2009}.

This circuit can be restructured into a pair of CV teleportation circuits~\cite{Braunstein1998} sandwiched between linear optics, as shown in Fig.~\ref{fig:macro}(b). To see this, note that equal phase delays commute past the
preceding 50:50 beamsplitter~%
\cite{Alexander2016}. Next, phase delays acting on modes $f(c)$ and $f(d)$ can be transferred to modes $f(c)^{\prime}$ and $f(d)^{\prime}$, respectively, using the symmetries of a two-mode squeezed state~\cite{Korolkova2002, Gabay2016}. Finally, the displacements $\hat{D}_{1}$ and $\hat{D}_{2}$ are commuted past the Gaussian unitaries~$\hat{B}(\frac{\pi}{4})$ and~$\hat{R}(\xi_{\pm})$, resulting in a new set of displacements,
${\hat{D}_{3} = \op D(g\alpha_3)}$ and ${\hat{D}_{4} = \op D(g\alpha_4)}$, where ${\op D(\alpha) \coloneqq e^{\alpha \op a^\dag - \alpha^* \op a}}$, ${\alpha_3 \coloneqq i m_{f(c)}+m_{f(a)}}$, and ${\alpha_4 \coloneqq i m_{f(d)} + m_{f(b)}}$. The \emph{gain parameter} ${g>0}$ allows us to tailor the noise associated with the teleportation~\cite{Takeda2013}, as shown next.

The case ${g=1}$ corresponds to the original CV teleportation protocol~\cite{Furusawa1998,Braunstein1998}---i.e., an identity gate with additive Gaussian noise introduced into the evolution~\cite{Furusawa1998,Braunstein1998,Alexander2016, Alexander2014}. Since this type of noise involves photon creation, it is undesirable for linear optics. Gain tuning may also enable tailoring the noise for specific applications~\cite{Takeda2013} in more general CV cluster-state computations.

%

%
%
%
By setting ${g=\tanh r}$, the noise model becomes pure amplitude damping (photon loss)~\cite{Polkinghorne1999, Ralph1999, Hofmann2001, Takeda2013}, as shown in Fig.~\ref{fig:macro}(c), 
with efficiency ${\gamma=g^2 = \tanh^{2}r}$. Direct calculation using the symplectic representation (with loss modeled as a beamsplitter~\cite{Yuen1978}) shows that the same loss channel~$\mathcal L$ applied to each of $m$ modes commutes with arbitrary (lossless) linear optics on those modes. This shows the equivalence of~(c) and~(d) in Fig.~\ref{fig:macro}.

Since the green macronodes in Fig.~\ref{fig:QRLnetwork} are uniformly spaced, we can further commute all loss to the beginning of the entire network. Then, finite squeezing for a depth-$k$ MBLO circuit %
results in an effective loss channel with efficiency $\gamma_{\text{eff}} \coloneqq \gamma^{2(k+1)}=(\tanh{r})^{4(k+1)}$ applied to each input state before (or, equivalently, after) the implemented linear optics, which are now considered lossless. %

This conversion of squeezing into loss enables direct comparison with other quantum computing architectures. In particular, we can use $\gamma_{\text{eff}}$ to compare the squeezing demands of MBLO to actual losses in a physical interferometer. A recent experiment~\cite{Carolan2015} implemented a general 6-mode interferometer with 42\%~average insertion loss, corresponding to $\gamma_{\text{eff}} = 0.58$. Achieving the same performance in MBLO (in an otherwise lossless implementation with depth~${k=6}$) would require ${r \approx 2.32}$, corresponding to 20.1~dB of squeezing. (Note: $\#\text{dB} = 10 \log_{10} e^{2r} \approx 8.69\ r$). While this is experimentally demanding, it is within reach of near-term technology  
given that the state of the art is now 15~dB~\cite{Vahlbruch2016}. 
%

%

\emph{\textsc{BosonSampling} with MBLO.}---%
%
%
With frugal experimental requirements, \textsc{BosonSampling} efficiently samples from a distribution that is strongly believed to be computationally hard to simulate~\cite{Aaronson2011}, making it of great interest as a potential candidate for demonstrating the first post-classical quantum algorithm.

In MBLO-based \textsc{BosonSampling}, we inject (Fig.~\ref{fig:QRLgen}) $n$~single photons and ${m-n}$~vacuum states into an ${(m+1)\times(k+2)}$-macronode QRL%
. 
We choose ${m=n^2}$
in order to ensure collision-free output configurations~\cite{Aaronson2011} and ${k=m}$ so we can implement an arbitrary unitary with nearest-neighbor interactions (Fig.~\ref{fig:LOnetwork})~\cite{Clements2016}.
Alternatively, one could implement Gaussian (``scattershot'') \textsc{BosonSampling}%
~\cite{Lund2014,Aaronson2013} by attenuating the squeezing in some of the two-mode squeezed states at the beginning of the protocol and photodetecting half of each one. This would
nondeterministically project input states into either vacuum or single-photon states. 
After MBLO, the output is switched out of the QRL (see Fig.~\ref{fig:QRLgen}), appropriately displaced,
and 
measured via
photodetection, thereby sampling from a statistical distribution of photon-number configurations~\cite{Aaronson2011}. %

With amplitude damping inherent to MBLO (due to finite squeezing), sometimes ${<n}$ photons in total will be measured at the output. Such an instance is a failure, and we repeat the protocol until success. MBLO has efficiency $\gamma_{\text{eff}}$ for each mode, so the success probability of the device is~$\gamma_{\text{eff}}^{n}$ (the ${m-n}$ vacuum states are unchanged by loss), and it takes ${T\coloneqq \gamma_{\text{eff}}^{-n}}$~trials on average to yield a single successful measurement event.

For an efficient implementation, it is necessary that $T=T_\poly(n)$ is some fixed polynomial in~$n$. %
Clearly, if $r$ is held constant, then $T=(\coth r)^{4n(k+1)}$ grows exponentially in~$n$. We can reduce this scaling by allowing the squeezing parameter~$r$ to grow with~$n$, but for efficiency in the \emph{squeezing}~\cite{Liu2016}, this must scale at most logarithmically with~$n$ in order to ensure the average energy $E \propto \sinh^2 r = \mathcal O(e^{2r})$ is polynomial in~$n$. %

%

\begin{figure}
\includegraphics[width=1\columnwidth]{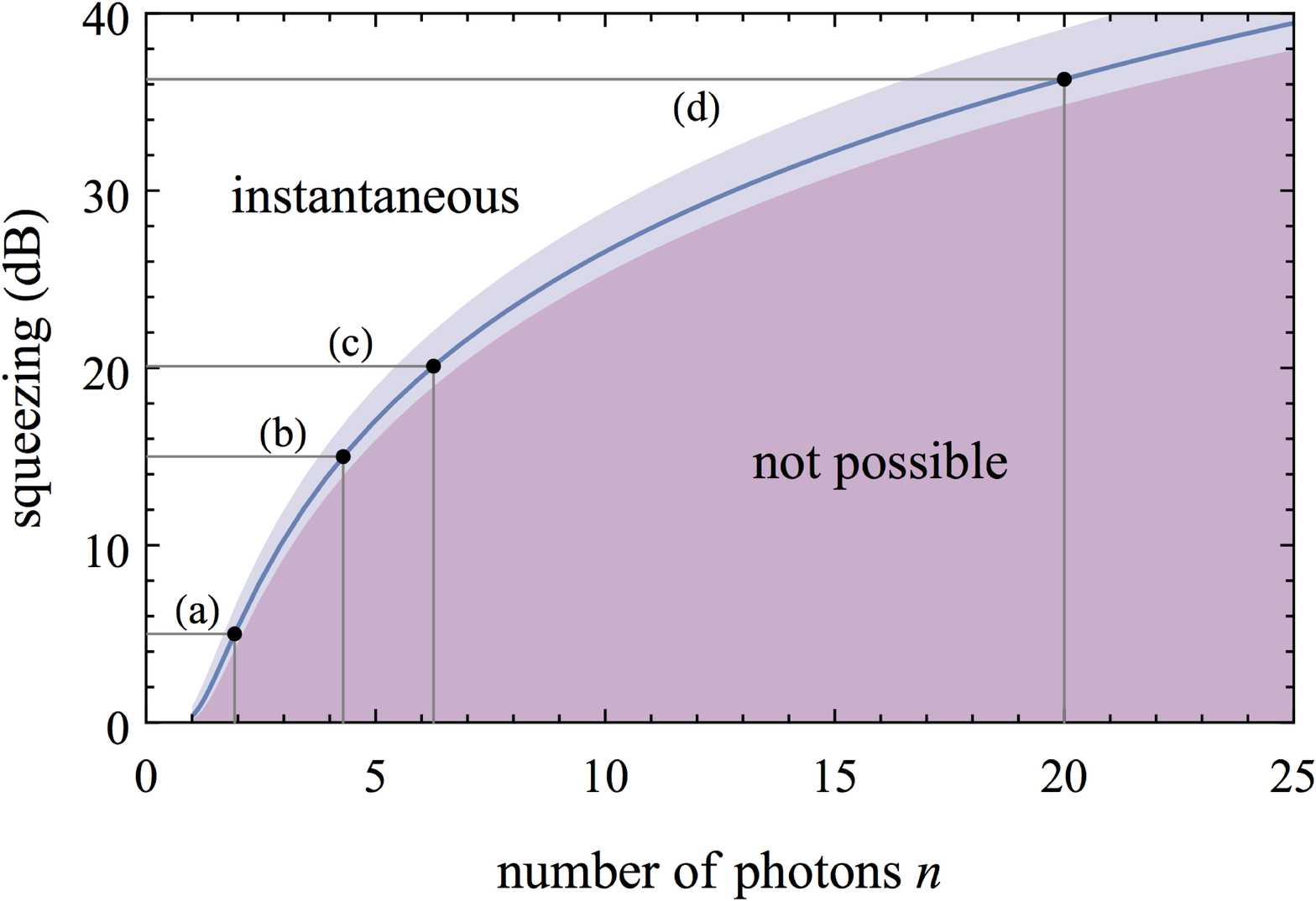}
\caption{Cluster-state squeezing levels currently required for MBLO-based \textsc{BosonSampling} of $n$ photons in ${m = n^2}$ modes using the temporal-mode implementation~\cite{Menicucci2011a,Yokoyama2013, Yoshikawa2016} and a simulated circuit of depth~${k=m}$ (see Fig.~\ref{fig:LOnetwork}). The blue line corresponds to experiments that would take 1~day on average for a successful run, while the surrounding blue region spans 1~minute (above) to 1~year (below). %
Marked points: (a)~5~dB, demonstrated in a large-scale cluster state~\cite{Yokoyama2013}, (b)~15~dB,  largest single-mode squeezing achieved in optics~\cite{Vahlbruch2016}; (c)~20.1~dB, squeezing corresponding to a recently reported 6-mode interferometer~\cite{Carolan2015}; (d)~36.3~dB, squeezing corresponding to ${n=20}$, a standard target~\cite{Aaronson2011} for \textsc{BosonSampling} to demonstrate an advantage over classical simulation.%
\label{fig:sampling}}
\end{figure}

Let $\ell \coloneqq 4n(k+1) = 4n(n^2+1)$, using $k=m=n^2$. Then, noting that the function $x \mapsto \tfrac 1 2 \log \coth x$ is self-inverse and that 
$1< \coth x < 1+x^{-1}$
for $x>0$, the relation $T = (\coth r)^\ell = T_\poly(n)$ implies
\begin{align}
\label{eq:rfromT}
	r &= \frac 1 2 \log \coth \left( \frac {\log T_\poly} {2\ell} \right) < \frac{1}{2} \log\left(1 + \frac{2 \ell}{\log T_\poly} \right),
\end{align}
ensuring that simultaneous efficiency in both time and squeezing is possible.

Having proven efficiency at the theoretical level, we now address practicality.
We assume the temporal-mode implementation (Fig.~\ref{fig:QRLgen}) with wave-packet duration $\Delta t\approx150 \text{ns}$~\cite{Yokoyama2013}%
, which means a single experiment requires $\tau\coloneqq(m+1)(k+2)\Delta t =(n^{2}+1)(n^{2}+2)\Delta t$ time to complete. Once $n$ photons have been successfully injected as inputs, a successful experiment will take time~$\tau T$ on average. In Fig.~\ref{fig:sampling}, using Eq.~\eqref{eq:rfromT}, we plot the squeezing required for $\tau T = \text{1~day}$ for various~$n$. The narrowness of the blue region ($1~\text{min} < \tau T < 1~\text{year}$) demonstrates that the blue line is effectively a hard boundary because of the exponential scaling with constant~$r$. In the lower (purple) region, greater than astronomical timescales would be required, while experiments in the upper (white) region are split second.

Notice that 
squeezing levels~$\sim$17~dB (only 
2~dB higher than current levels~\cite{Vahlbruch2016}) would enable sampling 5 photons from 25 modes, which would outperform the largest experimental demonstration to date (4 photons from 8 modes)~\cite{He2016}. Finally, the recently reported 6-mode interferometer~\cite{Carolan2015} has efficiency corresponding to 20.1~dB of squeezing (as noted above), which 
would
enable MBLO sampling 
of 6~photons from 36~modes---a much larger interferometer.

Our results are optimistic in that we have neglected additional noise sources (discussed below). On the other hand, they are also conservative in that we postselect on no lost photons for \textsc{BosonSampling}. Tolerating some loss through \emph{approximate} \textsc{BosonSampling}~\cite{Rahimi-Keshari2016} will likely allow for even lower squeezing while retaining computational hardness~\cite{Aaronson2016}.
%
%

\emph{Conclusion.}---%
%
Measurement-based linear optics offers a novel approach %
to large-scale interferometry. With MBLO, we get a large \emph{virtual} interferometer from a compact \emph{physical} setup. This compactness will be an advantage when experimentally confronting the usual sources of noise---e.g., mode mismatch~\cite{Rahimi-Keshari2016}, coupling losses~\cite{Yokoyama2013}, and phase locking~\cite{Yoshikawa2016}.

One might question the wisdom of employing squeezing (a nonlinear operation) for linear optics. For small experiments, this would be a valid objection. The value of MBLO lies in its simplicity and flexibility when implementing large-scale interferometers. %
Spurred on by the recent detection of gravitational waves~\cite{Abbott2016}, there is significant ongoing experimental drive to improve squeezing technology~\cite{Eberle2010,Mehmet2011,Aasi2013} for next-generation gravitational-wave astronomy~\cite{Caves1981}.
Experimental squeezing levels have recently been elevated to 15~dB~\cite{Vahlbruch2016} with high homodyne efficiency (99.5\%), high phase sensitivity ($1.7$~mrad), and low total optical losses (2.5\%). Progress in improving squeezing---a physical and information-theoretic resource~\cite{Braunstein2005a,Liu2016}---will directly translate into payoffs for other squeezing-based applications~\cite{Menicucci2014}, including MBLO.
Still, the relative benefit of MBLO over more conventional implementations of large-scale interferometry (e.g.,~interfering hundreds or thousands of modes) remains an important open question that we leave to future work.

\emph{Acknowledgments.}---%
We thank Carl Caves, Giulia Ferrini, Akira Furusawa, Nana Liu, Olivier Pfister, and Peter van~Loock for discussions. P.P.R.\ acknowledges support from Lockheed Martin. P.P.R.\ is funded by an ARC Future Fellowship (project FT160100397). This work was supported by the Australian Research Council~(ARC) under grant No.~DE120102204 and by the ARC Centre of Excellence for Quantum Computation and Communication Technology (CQC$^2$T), grant No.~CE170100012. Additional funding was provided by the U.S.\ Defense Advanced Research Projects Agency~(DARPA) Quiness program.
%



\end{document}